\documentclass[prd,showpacs,showkeys,preprint]{revtex4}

\usepackage{amsmath}
\usepackage{amssymb}
\usepackage{yfonts}[1988/10/03]
\usepackage{graphicx}
\newcommand {\beq}{\begin{equation}}
\newcommand {\eeq}{\end{equation}}
\newcommand {\beqa}{\begin{eqnarray}}
\newcommand {\eeqa}{\end{eqnarray}}
\newcommand {\n}{\nonumber \\}

\begin{document}
\title{ Gravitational Waves in a Spatially Closed de-Sitter Spacetime } 
\author{Amir H. Abbassi}
\email{ahabbasi@modares.ac.ir}
\author{Jafar Khodagholizadeh}
\email{j.gholizadeh@modares.ac.ir}
\affiliation{Department of Physics, School of Sciences, Tarbiat Modares University, P.O. Box 14155-4838, Tehran, Iran.}
\author{Amir M. Abbassi}
\email{amabasi@khayam.ut.ac.ir} 
\affiliation{ Department of Physics, University of Tehran, P.O. Box 14155-6455, Tehran, Iran.}

\begin{abstract}
 Perturbation of gravitational fields may be decomposed into scalar, vector and tensor components. In this paper we concern with the evolution of tensor mode perturbations in a spatially closed de-Sitter background of Robertson-Walker form. It may be thought as gravitational waves in a classical description. The chosen background has the advantage of to be maximally extended and symmetric. The spatially flat models commonly emerge from inflationary scenarios are not completely extended. We first derive the general weak field equations. Then the form of the field equations in two special cases, planar and spherical waves, are obtained and their solutions are presented. The radiation field from an isolated source is calculated. We conclude with discussing the significance of the results and their implications. 
   
\keywords{de-Sitter space, Gravitational waves, Tensor mode}

\end{abstract}
%\pacs{98.80.-k, 04.20.Cv, 02.40.-k}
%\preprint{***}
\maketitle

\section{Introduction}
Here we first investigate the freely propagating gravitational field requiring no local sources for their existence in a
particular background. As an essential feature of the analysis of general theory of small fluctuations, we assume that all departures from homogeneity and isotropy are small, so that they can be treated as first order perturbations. 
We focus our analysis on an unperturbed metric that has maximal extension and symmetry by taking $K=1$ and presence of a positive cosmological constant. The background is de-Sitter spacetime in slicing such that the spatial 
section is a 3-sphere. In the previous works mostly the case $K=0$ wre investigated extensively [1-5]. Even though 
in some works $K$ is not fixed for demonstrating the general field equations, but for solving them usually $K=0$ is imposed [6,7]. The study of this particular problem is interesting and relevant to present day cosmology for the following.
Seven-year data from WMAP with imposed astrophysical data put constraints on the basic parameters of cosmological
models. The dark energy equation of state parameter is $-1.1\pm 0.14$, consistent with the cosmological constant value of $-1$. While WMAP data alone can not constraint the spatial curvature parameter of the observable universe
$\Omega_k$ very well, combining the WMAP data with other distance indicators such as $H_0$ , BAO, or supernovae can constraint $\Omega_k$. Assuming $ \omega=-1$, we find  %-0.0133<\Omega_k<0.0084$, 
$\Omega_\Lambda=0.73 +/- 0.04$ and $\Omega_{total}=1.02 +/- 0.02$. Even though in WMAP seven-year data it has been concluded as an evidence in the support of flat universe, but in no way the data does not role out the case of $K=1$ [8]. In the nine-year data from WMAP the reported limit on spatial curvature parameter is $ \Omega_{k}=-0.0027^{+0.0039}_{-0.0038}$[9]. There is much hope Planck data reports release makes the situation more promising to settle down this dispute. But Planck 2013 results $ XXVI $,merely find no evidence for a multiply-connected topology with a fundamental domain within the last scattering surface. Further Planck measurement of CMB polarization probably provide more definitive conclusions[10]  . In the analysis of gravitational waves commonly
Minkowski metric is taken as the unperturbed background. According to mentioned observational data, the universe 
is cosmological constant dominated at our era. So in the analysis of gravitational waves we should replace the Minkowski background with de-Sitter metric. The essential point is that spatially open and flat de-Sitter spacetime are subspaces of spatially closed de-Sitter space. The first two are geodesically incomplete while the third is geodesically complete and maximally extended. From the singularity point of view the issue of completeness is crucial for a spacetime to be non-singular.  Taking the issue of completeness seriously, we have no way except to choose $K=1$. By choosing the maximally extended de-Sitter metric as our unperturbed background we include both cosmological and curvature terms in discussion of gravitaional waves [11-15]. We begin by deriving the required linear field equations. Then the solution of the obtained equation are discussed. At the end an attempt is done to solve the field equation by source. 
\section{ Linear Weak Field Equations}
 \label{}
  Supposed unperturbed metric components in Cartesian coordinate system are \cite{16}:
  \beqa
  \bar{g}_{00}&=&-1,\;\;\;\;\bar{g}_{i0}=0,\;\;\;\bar{g}_{ij}=a^2(t)\tilde{g}_{ij}\n
 a(t)&=&\alpha\cosh(t/\alpha),\;\;\;\tilde{g}_{ij}=\delta_{ij}+K\frac{x^ix^j}{1-Kx^2},
   \eeqa
 with the inverse metric
 \beqa
 \bar{g}^{00}=-1,\;\;\;\bar{g}^{0i}=0,\;\;\;\bar{g}^{ij}=a^2(t)\tilde{g}^{ij},\;\;\;\tilde{g}^{ij}=(\delta^{ij}-Kx^ix^j),
 \eeqa
 where $K$ is curvature constant and $\alpha=\sqrt{\frac3{\Lambda}}$.
 The non-zero components of the metric compatible connections are:
 \beqa
 \bar{\Gamma}^0_{ij}&=&a\dot{a}(\delta_{ij}+K\frac{x^ix^j}{1-Kx^2})=a\dot{a}\tilde{g}_{ij}\n
 \bar{\Gamma}^i_{0j}&=&\frac{\dot{a}}{a}\delta_{ij}\n
 \bar{\Gamma}^i_{jk}&=&\tilde{\Gamma}^i_{jk}=K\tilde{g}_{jk}x^i.
 \eeqa
 Dot stands for derivative with respect to time. Since we are working in a holonomic basis, then the connection is torsion-free or symmetric with respect to lower indices. Let us decompose the perturbed metric as:
 \beq
 g_{\mu\nu}=\bar{g}_{\mu\nu}+h_{\mu\nu},
 \eeq
 where $\bar{g}_{\mu\nu}$ is defined by eq.(1) and $h_{\mu\nu}$ is small symmetric perturbation term. The inverse metric is perturbed by
 \beq
 \chi^{\mu\nu}=g^{\mu\nu}-\bar{g}^{\mu\nu}=-\bar{g}^{\mu\rho}\bar{g}^{\nu\sigma}h_{\rho\sigma},
 \eeq
 with components
 \beqa
 \chi^{00}&=&-h_{00},\;\;\;\chi^{i0}=a^2(h_{i0}-Kx^ix^jh_{j0})\n
 \chi^{ij}&=&-a^{-4}(h_{ij}-Kx^ix^kh_{kj}-Kx^jx^kh_{ki}+K^2x^ix^jx^kx^lh_{kl}).
 \eeqa
 Perturbation of the metric produces a perturbation to the affine connection \cite{15}
 \beq
 \delta\Gamma^\mu_{\nu\lambda}=\frac12\bar{g}^{\mu\rho}[-2h_{\rho\sigma}\bar{\Gamma}^\sigma_{\nu\lambda}+\partial_\lambda h_{\rho\nu}+\partial_\nu h_{\rho\lambda}-\partial_\rho h_{\lambda\nu}].
 \eeq
 Thus eq.(7) gives the components of the perturbed affine connection as: 
 \beqa
 \delta\Gamma^i_{jk}&=&\frac1{2a^2}[-2a\dot{a}(h_{i0}-Kx^ix^l h_{l0})(\delta_{jk}+K\frac{x^kx^j}{1-Kx^2})
 -2K(h_{im}-Kx^ix^l h_{lm})(\delta_{jk}+K\frac{x^jx^k}{1-Kx^2})x^m\n
 &\;&+\partial_k h_{ij}+\partial_jh_{ik}-\partial_ih_{jk}
 -Kx^ix^j(\partial_k h_{lj}+\partial_j h_{lk}-\partial_l h_{jk})]\\
 \delta\Gamma^i_{j0}&=&\frac1{2a^2}(-2\frac{\dot{a}}{a}h_{ij}+\dot{h}_{ij}+\partial_j h_{i0}-\partial_ih_{j0}+2Kx^ix^k\frac{\dot{a}}{a}h_{kj}
 -Kx^ix^k\dot{h}_{kj}-Kx^ix^k\partial_j h_{k0}\n
 &\;&+Kx^ix^k\partial_k h_{j0})\\
 \delta\Gamma^0_{ij}&=&\frac12 [2a\dot{a}(\delta_{ij}+K\frac{x^ix^j}{1-Kx^2})h_{00}+2Kh_{0k}(\delta_{ij}x^k+K\frac{x^ix^jx^k}{1-Kx^2})
 -\partial_i h_{0j}-\partial_j h_{0i}+\dot{h}_{ij}]\\
 \delta\Gamma^i_{00}&=&\frac1{2a^2}[2\dot{h}_{0i}-\partial_i h_{00}-2Kx^ix^j\dot{h}_{0j}+Kx^ix^j\partial_j h_{00}]\\
 \delta\Gamma^0_{i0}&=&\frac{\dot{a}}{a} h_{i0}-\frac12\partial_i h_{00}\\
 \delta\Gamma^0_{00}&=&-\frac12\dot{h}_{00}.
 \eeqa
 The tensor mode perturbation to the metric can be put in the form
 \beq
 h_{00}=0,\;\;\; h_{i0}=0,\;\;\; h_{ij}=a^2D_{ij},
 \eeq
 where $D_{ij}$s are functions of $\vec{X}$ and $t$, satisfying the conditions
 \beq
 \tilde{g}^{ij}D_{ij}=0,\;\;\; \tilde{g}^{ij}\bar{\nabla}_iD_{jk}=0.
 \eeq
 The perturbation to the affine connection in tensor mode are:
 \beqa
 \delta\Gamma^0_{00}&=& \delta\Gamma^0_{i0}=\delta\Gamma^i_{00}=0\\
 \delta\Gamma^0_{ij}&=&a\dot{a}D_{ij}+\frac{a^2}{2}\dot{D}_{ij}\\
 \delta\Gamma^i_{j0}&=&a\dot{a}D_{ij}+\frac{a^2}{2}\dot{D}_{ij}\\
 \delta\Gamma^i_{jk}&=&\frac12[\partial_kD_{ij}+\partial_jD_{ik}-\partial_iD_{jk}-2K(D_{im}-Kx^ix^lD_{lm})\times\n
 &\;&(\delta_{jk}+K\frac{x^jx^k}{1-Kx^2})x^m-Kx^ix^l(\partial_kD_{lj}+\partial_jD_{lk}-\partial_lD_{jk})].
 \eeqa
 The Einstein field equation without matter source for the tensor mode of perturbation gives
 \beq
 \delta R_{jk}=-\Lambda a^2D_{jk},
 \eeq
 where
 \beqa
 \delta R_{jk}&=&-(2\dot{a}^2+a\ddot{a})D_{jk}-\frac32\dot{D}_{jk}-\frac{\dot{a}^2}{2}\ddot{D}_{jk}\n
 &\;&+\frac12\partial^i\partial_iD_{jk}-4KD_{jk}-\frac K2(\partial_i\partial_mD_{jk})x^ix^m\n
 &\;&-\frac32Kx^m\partial_mD_{jk}-K(\partial_kD_{mj}+\partial_jD_{mk})x^m\n
 &\;&+K^2D_{ml}(\delta_{jk}+K\frac{x^ix^k}{1-Kx^2})x^mx^l.
 \eeqa
 Scale factor $a(t)$ satisfies the Friedmann equation, so we get
 \beq
 2\dot{a}^2+a\ddot{a}=\Lambda a^2-2K.
 \eeq
 Inserting eq.(22) in eq.(21) and eq.(21) in eq.(20) , we would have
 \beqa
 &\;&-\frac32 a \dot{a}\dot{D}_{jk}-\frac{a^2}{2}\ddot{D}_{jk}-2KD_{jk}+\frac12 \partial^i\partial_i D_{jk}-\frac K2(\partial_i\partial_mD_{jk})x^ix^m-\frac32Kx^m\partial_mD_{jk}\n
 &\;&-K(\partial_kD_{mj}+\partial_jD_{mk})x^m+KD_{ml}(\delta_{jk}+K\frac{x^jx^k}{1-Kx^2})x^mx^l=0
 \eeqa
 It is straightforward to show that
 \beqa
 &\;&\frac12\nabla^2D_{jk}\equiv \frac12\bar{g}^{mn}\nabla_m\nabla_nD_{jk}=\frac12\partial_i\partial^iD_{jk}-\frac K2(\partial^i\partial_mD_{jk})x^ix^m-KD_{jk}\n
 &\;&-Kx^l(\partial_kD_{lj}+\partial_jD_{lk})-\frac32Kx^i\partial_iD_{jk}+Kx^lx^i(\delta_{jk}+K\frac{x^jx^k}{1-Kx^2})D_{li}.
 \eeqa
 It remains to put eq.(24) in eq.(23), then we obtain the final equation.
 \beq
 \frac12\nabla^2D_{jk}-\frac32a\dot{a}\dot{D}_{jk}-\frac{a^2}{2}\ddot{D}_{jk}-KD_{jk}=0.
 \eeq
 Our first task to establish the field equations is fulfilled. Next we look for special solutions of this field equation analogue to plane and spherical waves. For the plane wave like solutions, using Cartesian coordinate systems is suitable while for the spherical waves, polar coordinates $(\chi,\theta,\phi)$ are convenient.

\section{Plane-wave analogue}
\label{}
In the case of flat models i.e. $K=0$ condition (15) reduces to
\beq
D_{ii}=0,\;\;\;\delta^{ik}\partial_iD_{kj}=0.
\eeq
Looking for a wave  propagating in z-direction,  eq.(26) simply gives
\beq
D_{i3}=0,
\eeq
with two independent modes
\beq
D_{+ij}=D_+(z,t)\left(\begin{array}{ccc} 1&0&0\cr 0 &-1 &0 \cr 0&0&0\cr \end{array}\right),\;\;\;and
\;\;\;D_{\times ij}=D_\times(z,t)\left(\begin{array}{ccc}
0&1&0\cr 1&0&0\cr 0&0&0\cr\end{array}\right),
\eeq
where $D(z,t)$ satisfies $\Box ^2D(z,t)=0$ with the well-known plane wave solution.\\
In the case of $K=1$ an analogue solution for eq.(15) exists. Following a lengthy calculation due to non-diagonal components of $\tilde{g}_{ij}$ we obtain
\beq
D_{+ij}=\frac{D_+(z,t)}{\sqrt{1-X^2}}\left(\begin{array}{ccc}
\frac1{1-x^2-z^2}&0&\frac{xz}{(1-z^2)(1-x^2-z^2)}\cr
0&-\frac1{1-y^2-z^2}&\frac{-yz}{(1-z^2)(1-y^2-z^2)}\cr
\frac{xz}{(1-z^2)(1-x^2-z^2)}&\frac{-yz}{(1-z^2)(1-y^2-z^2)}&\frac{z^2(x^2-y^2)}{(1-z^2)(1-x^2-z^2)(1-y^2-z^2)}\cr\end{array}\right),
\eeq
where $X^2=x^2+y^2+z^2$ and
\beq
D_{\times ij}=\frac{D_\times (z,t)}{\sqrt{1-X^2}(1-y^2-z^2)}\left(\begin{array}{ccc}
0&1&\frac{yz}{1-z^2}\cr 1&\frac{2xy}{1-y^2-z^2}&\frac{xz(1+y^2-z^2)}{(1-z^2)(1-y^2-z^2)}\cr \frac{yz}{1-z^2}&\frac{xz(1+y^2-z^2)}{(1-z^2)(1-y^2-z^2)}&\frac{2xyz^2}{(1-z^2)(1-y^2-z^2)}\end{array}\right).
\eeq
By inserting eqs.(29) and (30) in eq.(25) with some manipulation we conclude that each mode, $\times$ and $+$, satisfies
\beqa
&\;&(1-z^2)\frac{\partial^2}{\partial z^2}D(z,t)+3z\frac{\partial}{\partial z}D(z,t)-D(z,t)+\frac{6D(z,t)}{1-z^2}\n
&\;&-3a\dot{a}\dot{D}(z,t)-a^2\ddot{D}(z,t)-2D(z,t)=0.
\eeqa
We use method of separation of variables to find the solutions of eq.(31). Then we may write
\beq
D(z,t)=D(z)\hat{D}(t).
\eeq
Using eq.(32), eq.(31) leads to
\beqa
&\;&\frac{1-z^2}{D(z)}\frac{\partial^2}{\partial z^2}D(z)+\frac{3z}{D(z)}\frac{\partial}{\partial z}D(z) -1+\frac6{1-z^2}\n
&\;&= a^2\frac{\ddot{\hat D}(t)}{\hat{D}(t)}+3a\dot{a}\frac{\dot{\hat D}(t)}{\hat{D}(t)}+2.
\eeqa
Eq.(33) may hold merely if each side is equal to a constant, i.e. we have:
\beqa
&\;&\frac{1-z^2}{D_q(z)}\frac{\partial^2}{\partial z^2}D_q(z)+\frac{3z}{D_q(z)}\frac{\partial}{\partial z}D_q(z)-1+\frac6{1-z^2}=-q^2,\\
&\;&\frac{a^2\ddot{\hat D}(t)}{\hat{D}_q(t)}+\frac{3a\dot{a}}{\hat{D}_q(t)}\dot{\hat D}_q(t)+2=-q^2,
\eeqa
where $q^2$ is an arbitrary positive constant. We should take it positive since we are looking for a periodic wave. Eqs.(34) and (35) can be written as
\beq
(1-z^2)\frac{\partial ^2}{\partial z^2}D_q(z)+3z\frac{\partial}{\partial z}D_q(z)+(q^2-1+\frac6{1-z^2})D_q(z)=0,
\eeq
and
\beq
a^2\ddot{\hat D}_q(t)+3a\dot{a}\dot{\hat D}(t)+(q^2+2)\hat D(t)=0.
\eeq
To solve eq.(36) and finding $D_q(z)$ we define
\beq
D_q(z)=(1-z^2) U_q(z),
\eeq
inserting eq.(38) in eq.(36) we get the following equation for$ U_q(z) $
\beq
(1-z^2)\frac{d^2}{dz^2}U_q(z)-z\frac{d}{dz}U_q(z)+(q^2+3)U_q(z)=0
\eeq
We notice that the solutions of eq.(39) may be written as Chebyshev polynominal of type I provided we take, $ q^{2}=n^{2}-3 $, where $ n $ is integer and $ U_{n}(z) $ is
\beq
U_{n}(z)\propto exp(\pm in \arccos z)
\eeq
So we have
\beq
D_{n}(z)=(1-z^2) exp(\pm i n \arccos z),
\eeq
and
\beq
\int_{-1}^{1}(1-z^{2})^{\dfrac{-5}{2}} D_{n}(z)D_{n^{'}}^{\ast} dz
=\dfrac{1}{\pi}\delta_{n n^{'}}.
\eeq
Next we examine the temporal dependence of this mode. Putting eq.(41) in eq.(37) gives
\beq
a^2\ddot{\hat D}_n(t)+3a\dot{a}\dot{\hat D}_n(t)+(n^2-1)\hat{D}_n(t)=0.
\eeq
It is convenient to define the conformal time $\tau$ by:
\beq
d\tau=\frac{dt}{a(t)}\;\;\;\hbox{where}\;\;\;a(t)=\alpha\cosh(t/\alpha)
\eeq 
Integrating eq.(44) gives:
\beq
exp(t/\alpha)=tan(\tau/2).
\eeq
Notice that $t=-\infty, 0,+\infty$ corresponds to $\tau=0,\pi/2,\pi$ respectively. So while the domain of  coordinate time is $-\infty<t<+\infty$, the domain of conformal time is $0<\tau<\pi$.\\
We may recast eq.( 43) in terms of conformal time as
\beq
\frac{d^2}{d\tau^2}\breve{D}_n(\tau)-2\cot\tau\frac d{d\tau}\breve{D}_n(\tau)
+(n^2-1)\breve{D}_n(\tau)=0,
\eeq
where $\hat{D}_n(t)=\breve{D}_n(\tau)$. To solve eq.(46), let us define a new parameter $Y=\cos\tau$ with domain $-1<Y<+1$ and $t=-\infty,0,+\infty$ correspond to $Y=1,0,-1$ respectively. In terms of the new parameter $Y$, eq.(46) becomes
\beq
(1-Y^2)\frac{d^2}{dY^2}\tilde{D}_n(Y)+Y\frac{d}{dY}\tilde{D}_n(Y)+(n^2-1)\tilde{D}_n(Y)=0,
\eeq
where $\breve{D}_n(\tau)=\tilde{D}_n(Y)$.\\
If we define $W_n(Y)=\frac{d}{dY}\tilde{D}_n(Y)$ and differentiate eq.(47) with respect to $Y$, this gives
\beq
(1-Y^2)\frac{d^2}{dY^2}W_n(Y)-Y\frac{d}{dY}W_n(Y)+n^2W_n(Y)=0.
\eeq
Again eq.(48) is Chebyshef of first kind and its solutions are:
\beq
W_n(Y)=exp(\pm in\arccos Y).
\eeq
For last step we should solve
\beq
\frac{d}{dY}\tilde{D}_n(Y)=exp(\pm in\arccos Y).
\eeq
We have previously defined $Y=\cos\tau$, so we have $\tau=\arccos Y$. Let us recast eq.(50) in terms of $\tau$, it becomes
\beq
\frac{d}{d\tau}\breve{D}_n(\tau)=-\sin\tau e^{\pm in\tau},
\eeq
with
\beq
\breve{D}_n(\tau)=\frac1{1-n^2}(\cos\tau\mp in\sin\tau)e^{\pm in\tau}.
\eeq 
We may write
\beq
\tilde{D}_n(z,\tau)=\frac{(1-z^2)}{1-n^2}(\cos\tau \mp in\sin\tau)\left\{
\begin{array}{c}
exp[\pm in(\arccos z+\tau)]\cr
exp[\pm in (\arccos z-\tau)].
\end{array}\right.
\eeq
The first mode $n=1$ is pure gauge mode and should be excluded from the acceptable solutions.
This is the analogue of a plane wave moving in z-direction for a closed model. We may find a similar solution for the waves that are analogue to plane wave in x and y directions. In this case we would have:
\beqa
D_{+ij}(x,y,z,t)&=&\frac{D_+(x,t)}{\sqrt{1-X^2}}\left(\begin{array}{ccc}
\frac{x^2(z^2-y^2)}{(1-x^2)(1-x^2-y^2)(1-x^2-z^2)}&\frac{-xy}{(1-x^2)(1-x^2-z^2)}&\frac{xz}{(1-x^2)(1-x^2-z^2)}\cr
\frac{-xy}{(1-x^2)(1-x^2-z^2)}&\frac{-1}{(1-x^2-y^2)}&0\cr
\frac{xz}{(1-x^2)(1-x^2-z^2)}&0&\frac1{(1-x^2-z^2)}\cr\end{array}\right),\\
D_{\times ij}(x,y,z,t)&=&\frac{D_{\times}(x,t)}{(1-x^2-y^2)\sqrt{1-X^2}}\left( \begin{array}{ccc}
\frac{2x^2yz}{(1-x^2)(1-x^2-y^2)}&\frac{xz(1-x^2+y^2)}{(1-x^2)(1-x^2-y^2)}&\frac{xy}{1-x^2}\cr
\frac{xz(1-x^2+y^2)}{(1-x^2)(1-x^2-y^2)}&\frac{2yz}{(1-x^2-y^2)}&1\cr
\frac{xy}{(1-x^2)}&1&0\end{array}\right),\\
D_{+ij}(x,y,z,t)&=&\frac{D_+(y,t)}{\sqrt{1-X^2}}\left(\begin{array}{ccc}
\frac1{1-x^2-y^2}&\frac{xy}{(1-y^2)(1-x^2-y^2)}&0\cr
\frac{xy}{(1-y^2)(1-y^2-z^2)}&\frac{y^2(x^2-z^2)}{(1-y^2)(1-y^2-z^2)(1-x^2-y^2)}&\frac{-yz}{(1-y^2)(1-y^2-z^2)}\cr
0&\frac{-yz}{(1-y^2)(1-y^2-z^2)}&\frac{-1}{1-y^2-z^2}\end{array}\right),\\
D_{\times ij}(x,y,z,t)&=&\frac{D_{\times}(y,t)}{(1-y^2-z^2)\sqrt{1-X^2}}\left(\begin{array}{ccc}
0&\frac{yz}{1-y^2}&1\cr
\frac{yz}{1-y^2}&\frac{2xy^2z}{(1-y^2)(1-y^2-z^2)}&\frac{xy(1-y^2-z^2)}{(1-y^2)(1-y^2-z^2)}\cr
1&\frac{xy(1-y^2+z^2)}{(1-y^2)(1-y^2-z^2)}&\frac{2xz}{1-y^2-z^2}\end{array}\right),
\eeqa
where $D(x,t)$ and $D(y,t)$ are given by:
\beq
\hat{D}_n(x,\tau)=\frac{1-x^2}{1-n^2}(\cos\tau\mp i n \sin\tau)\left\{
\begin{array}{c}
exp[\pm i n (\arccos x+\tau)]\cr 
exp[\pm i n (\arccos x-\tau)]\cr
\end{array}\right.  n\neq 1
\eeq
and
\beq
\hat{D}_n(y,t)=\frac{1-y^2}{1-n^2}(\cos\tau\mp i n\sin\tau)\left\{\begin{array}{c}
exp[\pm i n(\arccos y+\tau)]\cr
exp[\pm in(\arccos y-\tau)]\end{array}\right.  n\neq 1.
\eeq
It is important to notice that eq.(58) and Eq.(59) may be achieved  from eq.(29) and eq.(30)  respectively by 
the coordinate transformation $x\to z,\;\;y\to y , \;\; $ and $z\to -x$. This leads us to write the solution of waves moving in an arbitrary direction. Let us assume this arbitrary direction is 
\beq
\hat{n}_1=\sin\theta\sin\varphi , \;\;\;\hat{n}_2=\sin\theta\cos\varphi , \;\;\;\hat{n}_3=\cos\theta.
\eeq
The result can be obtained from eqs.(29) and (30) by the coordinate transformation introduced by
\beqa
z &\to& \hat{n}\cdot\vec{X}\n
y&\to&\frac{\hat{n}_3(\hat{n}\cdot\vec{X})-z}{\sqrt{1-\hat{n}_3^2}}\n
x&\to&\frac{\hat{n}_2 x-\hat{n}_1 y}{\sqrt{1-\hat{n}_3^2}}
\eeqa
We get
\beq
{{D_n}_{\buildrel  +\over \times}}_{ij}(\vec{X},t)={A_{\buildrel +\over \times}}_{ij}(\vec{X},\hat{n})
\frac{(1-(\hat{n}\cdot\vec{X})^2)}{1-n^2}(\cos\tau\mp i n \sin\tau)\times\left\{\begin{array}{c}
exp[\pm i n (\arccos(\hat{n}\cdot\vec{X})+\tau]\cr
exp[\pm i n (\arccos(\hat{n}\cdot\vec{X})-\tau]\end{array}\right.  \;\;n\neq 1,
\eeq
where the explicit forms of ${A_{\buildrel +\over \times}}_{ij}(\vec{X},\hat{n})$ are listed in the appendix.
This result may be used to expand a general function as linear superposition of these eigenfunctions, i.e. it should be replaced with $exp(ik_\mu x^\mu) $ in Fourier transformations.   
\section{spherical wave analogue}
To consider this case it is more suitable to work in polar coordinates, $x^i=(\chi,\theta,\phi)$. In this basis the non-zero components of the unperturbed metric are:
\beq
\tilde{g}_{11}=1,\;\;\;\tilde{g}_{22}=\sin^2\chi,\;\;\;\tilde{g}_{33}=\sin^2\chi\sin^2\theta,
\eeq
with the inverse
\beq
\tilde{g}^{11}=1,\;\;\;\tilde{g}^{22}=\sin^{-2}\chi,\;\;\;\tilde{g}^{33}=\sin^{-2}\chi\sin^{-2}\theta.
\eeq
The non-zero components of the unperturbed connections are
\beq
\begin{array}{ll}
\Gamma^1_{22}=-\sin\chi\cos\chi,&\Gamma^1_{33}=-\sin\chi\cos\chi\sin^2\theta ,\cr
\Gamma^2_{21}=\cot\chi,&\Gamma^2_{33}=-\sin\theta\cos\theta ,\cr
\Gamma^3_{31}=\cot\chi,&\Gamma^3_{32}=\cot\theta .\cr\end{array}
\eeq
In this case $\tilde{g}_{ij}$ is diagonal and the conditions (15) for a transverse wave give
\beq
D_{1i}=0.
\eeq
We may distinguish two independent polarizations as
\beqa
D_{+ij}(\chi,\theta,t)&=&\frac{D_+(\chi,t)}{\sin^2\theta}\left(\begin{array}{ccc}
0&0&0\cr
0&1&0\cr
0&0&-\sin^2\theta \end{array}\right)\\
D_{\times ij}(\chi,\theta,t)&=&\frac{D_\times(\chi,t)}{\sin\theta}\left(\begin{array}{ccc}
0&0&0\cr
0&0&1\cr
0&1&0\end{array}\right).
\eeqa
Inserting eqs.(64) and (65) in eq.(25) and expressing $\nabla^2$ in polar coordinates, with a rather lengthy but straightforward calculation it can be shown that both $D_+(\chi,t)$ and $D_\times(\chi,t)$ must satisfy the same equation as \beq
\frac{\partial^2}{\partial \chi^2}D(\chi,t)-2\cot \chi\frac{\partial}{\partial \chi}D(\chi,t)+2\frac{D(\chi,t)}{\sin^2\chi}-3a\dot{a}\dot{D}(\chi,t)-a^2\ddot{D}(\chi,t)=0
\eeq
To solve eq.(69) we may assume that
\beq
D(\chi,t)=D(\chi)\hat{D}(t)
\eeq
Then we have
\beq
\frac1{D(\chi)}\frac{\partial^2}{\partial\chi^2}D(\chi)-2\frac{\cot\chi}{D(\chi)}\frac{\partial}{\partial\chi}D(\chi)+\frac2{\sin^2\chi}=a^2\frac{\ddot{\hat D}(t)}{\hat{D}(t)}+3a\dot{a}\frac{\dot{\hat D}(t)}{\hat{D}(t)}.
\eeq
Eq.(71) holds provided that each side is equal to a constant, i.e.
\beqa
&\;&\frac{1}{D(\chi)}\frac{\partial^2}{\partial \chi^2}D(\chi)-2\frac{\cot\chi}{D(\chi)}\frac{\partial}{\partial\chi}D(\chi)+\frac2{\sin^2\chi}=-(q^2-1),\\
&\;&\frac{a^2(t)\ddot{\hat D}(t)}{\hat{D}(t)}+\frac{3a(t)\dot{a}(t)}{\hat{D}(t)}\dot{\hat D}(t)=-(q^2-1),
\eeqa
where $q^2$ is an arbitrary positive constant. So we have
\beqa
&\;&\frac{\partial^2}{\partial\chi^2}D(\chi)-2\cot\chi\frac{\partial}{\partial\chi}D(\chi)+(q^2+\frac2{\sin^2\chi})D(\chi)=0,\\
&\;&a^2(t)\ddot{\hat D}(t)+3a(t)\dot{a}(t)\dot{\hat D}+(q^2-1)\hat{D}(t)=0.
\eeqa
To solve eq.(74) for $D_q(\chi)$ we may define a new parameter $X=\cos\chi$ and $D(\chi)=\hat{D}(X)$, then eq.(74) gives: 
\beq
(1-X^2)\frac{d^2}{dX^2}\hat{D}(X)+X\frac{d}{dX}\hat{D}(X)+(q^2-1+\frac2{1-X^2})\hat{D}(X)=0.
\eeq
Eq.(76) has a solution as
\beq
\hat{D}_q(X)=(1-X^2)^{1/2}\frac{d}{dX}U_q(X),
\eeq
where $U_q(X)$ satisfy the following equation
\beq
(1-X^2)\frac{d^2}{dX^2}U_q(X)+X\frac{d}{dX}U_q(X)+(q^2-1)U_q(X)=0.
\eeq
Now we take $V_q(X)=\frac{d}{dX}U_q(X)$ which satisfies 
\beq
(1-X^2)\frac{d^2}{dX^2}V_q(X)-X\frac{d}{dX}V_q(X)+(q^2-1)V_q(X)=0
\eeq
Eq.(79) is a Chebyshef type I provided that we take $q=n$. Then we have
\beq
V_n(x)=exp(\pm in\arccos X),
\eeq
and
\beq
D_n(\chi)=\sin\chi exp(\pm in\chi).
\eeq
The temporal part is the same as plane wave analogue eq.(52) and we have
\beq
D_n(\chi ,t)=\frac{(\cos\tau\pm in\sin\tau)}{1-n^2}\sin\chi\left\{\begin{array}{c}
exp(\pm in (\chi+\tau))\cr
exp(\pm in(\chi-\tau)).\end{array}\right.
\eeq
It is interesting to note that in the case of flat models, i.e. $K=0$, eq.(74) in the $(r,\theta,\phi)$ bais takes the form
\beq
\frac{d^2}{dr^2}D(r)-\frac2r\frac{d}{dr}D(r)+\frac2{r^2}D(r)=-q^2D(r),
\eeq
which has the solution
\beq
D_q(r)\propto re^{\pm iqr},
\eeq
where $q$ can be any arbitrary real number. If we consider the ratio $\frac{h_{22}}{g_{22}}$ we get
\beqa
\frac{h_{22}}{g_{22}}&\propto &\frac1{\sin\chi}\left\{\begin{array}{c}
exp(\pm in(\chi+\tau))\cr
exp(\pm in(\chi-\tau))\cr\end{array}\right. \quad\hbox{for}\; K=1\\
\frac{h_{22}}{g_{22}}&\propto &\frac1r exp(\pm iqr), \quad\hbox{for}\; K=0.
\eeqa
Eqs.(85) and (86) both decrease by increasing the radial coordinate. The radiation fields(EM or GW)always decreases as inverse of radial coordinate regardless it is dipole or quadrupole field. its dependence on the source properly appears as a factor which could be dipole or quadrupole.
\section{The effect of gravitational waves}
To obtain a measure of the waves effect in this background ,we consider the rotation of the nearby particles as described by the geodesic deviation equation. For some nearly particles with four-velocity$ U^{\mu}(x) $ and sepration vector $ S^{\mu} $, we have
\begin{equation}
\dfrac{D^{2} S^{\mu}}{d\tau^{2}}= R^{\mu}_{~\nu \rho\sigma} U^{\mu} U^{\rho} S^{\sigma}.
\end{equation} 
Since the  Riemann tensor is already first order for test particles that are moving slowly we may write$ U^{\mu}=(1,0,0,0)$ in eq.(87).
So in computing eq.(87)we only need $ R^{i}_{~00j} $which is 
\begin{equation}
R^{i}_{~00j}=\dfrac{\ddot{a}}{a}\delta^{i}_{j}+\dfrac{1}{2} \tilde{g}^{ik} \ddot{D}_{kj}+\dfrac{\dot{a}}{a} \tilde{g}^{ik} \dot{D}_{kj}.
\end{equation}
For slowly-moving particles we have$ \tau =\dot{x}=t $ to lowest order so the geodesic deviation equation becomes
\begin{equation}
\dfrac{\partial ^{2}}{\partial t^{2}}S^{i}=R^{i}_{~00j} S^{j}.
\end{equation} 
The first term in eq.(88) will came a exponential expansion which is a general characteristic of de Sitter space .of course this is not the effect of gravitational waves and we may ignore it .
Making use of eq.(43) and eq.(52) and noticing that $ n\gg 1 $ .
The contribution of last term in eq.(89) vanishes, so we have: 
\begin{equation}
\dfrac{\partial^{2}}{\partial t^{2}}S^{i} =\dfrac{1}{2}\dfrac{\partial^{2}}{\partial t^{2}} D^{i}_{~j} S^{j}.
\end{equation}
To be specific  we chose eq.(29), as a wave moving in $ z-$direction , so we have 
\beq
D^{i}_{+j}=\frac{D_{+}(z,t)}{\sqrt{1-X^2}(1-z^2)}\left(\begin{array}{ccc}
1& \dfrac{xy}{1-y^{2}-z^{2}}&\frac{xz}{1-y^{2}-z^2}\cr \dfrac{-xy}{1-x^{2}-z^{2}}& -1& \dfrac{-yz}{1-x^{2}-z^{2}}\cr 0&0&0\end{array}\right).
\eeq
The eigenvalues of matrix eq.(91) are 
\begin{equation}
\lambda =0 ~~~~~,~~~~~ \lambda_{\pm}=\pm \dfrac{D_{+}(z,t)}{\sqrt{1-X^{2}}(1-z^{2})}\sqrt{1-\dfrac{x^{2}y^{2}}{(1-y^{2}-z^{2})(1-x^{2}-z^{2})}}
\end{equation}
Certainly the  $ z-$component of the separation vector is not affected by this gravitational waves.
As the case of flat space a circle of particles hit by a gravitational wave has a oscillatory motion but in closed spaces the normal axis of oscillation rotate at different location of spacetime.
\section{Generation of Gravitational Waves}
With presenting plane-wave solutions to the linearized vacuum field equation, it remains to discuss the generation of gravitational radiation by source. For this purpose it is necessary to consider the equation coupled to matter. Making use of completeness relation of the eigenfunction of vacuum equation we may write the solution of the field equation with source as 
\begin{eqnarray}
D_{ij}(\vec{x},\tau)=+16\pi G\sum_{n}\sum_{m}\int \dfrac{d^{2}\hat{n}}{4\pi} \int \dfrac{d^{3} x^{'} }{\sqrt{1-x^{'2}}}  \dfrac{d\tau^{'}}{\sin^2 \tau^{'}}  (1-(\hat{n}.\vec{x})^{2})(1-(\hat{n}.\vec{x^{'}})^{2})\nonumber \\ \times \exp[in(\arccos(\hat{n}.\vec{x})-\arccos(\hat{n}.\vec{x^{'}}))] (\cos\tau +im \sin\tau)\nonumber \\ \times(\cos \tau^{'}-im \sin \tau^{'}) \exp [-im(\tau -\tau^{'})]\dfrac{T_{ij}(\vec{x^{'}},\tau^{'})}{(1-m^{2})^{2}(n^{2}-m^{2})}
\end{eqnarray}
We assume the source is isolated , far away and slowly moving. This implies $\mid \vec{x^{'}}\mid \ll \mid \vec{x}\mid $. Also we take the distance to the source is not of cosmic scale so that$ \mid \vec{x}\mid \ll 1 $. With these approximations eq.(93)may be written as 
\begin{eqnarray}
D_{ij}(\vec{x},\tau)=+16\pi G\sum_{n}\sum_{M}\int \dfrac{d^{2}\hat{n}}{4\pi} \int d^{3} x^{'}  \dfrac{d\tau^{'}}{\sin^2\tau^{'}} \exp
[in(\vec{x}-\vec{x^{'}}).\hat{n}]\nonumber \\ \times [\cos(\tau -\tau^{'})+im\sin(\tau-\tau^{'})+(m^{2}-1)\sin\tau \sin \tau^{'}] \exp [-im(\tau -\tau^{'})]\dfrac{T^{ij}(\vec{x^{'}},\tau^{'})}{(1-m^{2})^{2}(n^{2}-m^{2})}.\n
\end{eqnarray}
Performing summation on $ m $ gives
\begin{eqnarray}
\sum_{m}\dfrac{\cos (\tau-\tau^{'})+im \sin (\tau-\tau^{'})+(m^{2}-1)\sin\tau \sin\tau^{'}}{(1-m^{2})^{2}(n^{2}-m^{2})} \exp(-im(\tau-\tau^{'}) \nonumber \\ =\dfrac{2\pi\theta(\tau-\tau^{'})}{(1-n^{2})^{2}n}[-\cos (\tau-\tau^{'}) \sin (n(\tau-\tau^{'}))+n \sin (\tau-\tau^{'}) \cos(n(\tau-\tau^{'}))+(n^{2}-1)\sin\tau \sin \tau^{'}].\n
\end{eqnarray}
Integrating on $ \hat{n} $ and putting eq.(95)in eq.(94)gives
\begin{eqnarray}
D_{ij}(\vec{x},\tau)=-\dfrac{16\pi G}{R}\sum_{n}\dfrac{\sin (nR)}{n^{2}(1-n^{2})} \int d\tau^{'}\theta (\tau-\tau^{'}) ~~~~~~~~~~~~~~~~~~~~~~~~~~~~~\nonumber \\ \times \dfrac{\cos (\tau-\tau^{'}) \sin(n(\tau-\tau^{'}))- n\sin (\tau-\tau^{'})\cos(n(\tau-\tau^{'}))+(n^{2}-1)\sin\tau \sin \tau^{'} }{\sin \tau^{'2}}\int d^{3}x^{'} T^{ij}(\vec{x^{'}},\tau^{'}).\n
\end{eqnarray}
Since the source is localized and spacetime locally looks flat , we have :
\beq
T^{\mu\nu}_{~,\nu}=0.
\eeq
Putting $ \mu=0 $ in eq.(97)and differentiating with respect to $ x^{0} $ gives:
\beq
T^{00}_{~,00}=-T^{0i}_{~,i0}.
\eeq
Then putting $ \mu=k $in eq.(97) gives:
\beq
T^{k0}_{~,0}+T^{ki}_{~,i}=0.
\eeq
Combining eq.(98) and eq.(99) leads to
\beq
T^{00}_{~,00}+T^{ik}_{~,ik}=0.
\eeq
Now we multiply both sides of eq.(100)by $ x^{n}x^{m} $ and integrate by parts two times over all space, ignoring the surface terms, finishing up with
\begin{eqnarray}
\int  T^{mn}(\vec{x},t)d^{3}x^{'}&=&\dfrac{1}{2}\dfrac{\partial^{2}}{\partial t^{2}} \int T^{00}x^{m}x^{n} d^{3}x \nonumber\\ &=&\dfrac{1}{2} \ddot{I}_{mn}(t),
\end{eqnarray}
where $ I_{mn} $ is the quadrupole moment of the mass distribution  of the source. Inserting eq.(101) in eq.(96) gives
\begin{eqnarray}
D_{ij}(\vec{x},\tau)=-\dfrac{8\pi G}{R}\sum_{n}\dfrac{\sin nR}{n^{2}(1-n^{2})} \int \dfrac{d\tau^{'}}{\sin^2\tau^{'}} \theta (\tau-\tau^{'}) ~~~~~~~~~~~~~~~~~~~~~~~\nonumber \\ \times [\cos (\tau-\tau^{'}) \sin(n(\tau-\tau^{'}))- n\sin (\tau-\tau^{'})\cos(n(\tau-\tau^{'}))+(n^{2}-1)\sin\tau \sin \tau^{'}] \ddot{I}_{ij}(\tau^{'})\n
\end{eqnarray}
The result of summation on $ n $ gives the retarded solution as:
\begin{eqnarray}
D_{ij}(\vec{x},\tau)=+\dfrac{8\pi G}{R} \int \dfrac{d\tau^{'}}{\sin^2\tau^{'}}\theta(\tau-\tau^{'})\theta(\tau-R-\tau^{'})~~~~~~~~~~~~~~~~\nonumber \\ \times \dfrac{2\pi i}{4}[+\dfrac{1}{2} \cos(\tau-\tau^{'}) \cos (\tau-\tau^{'}-R)+\dfrac{1}{2}\sin(\tau-\tau^{'}) \sin(\tau-\tau^{'}-R)]\ddot{I}_{ij}. 
\end{eqnarray}
Therefore we have: 
\begin{eqnarray}
D_{ij}(\vec{x},\tau)&=&\dfrac{2 \pi^{2} G}{R} i \int\dfrac{d\tau^{'}}{\sin^2\tau^{'}}\ddot{I}_{ij}\theta(\tau -\tau^{'})
\theta(\tau-R-\tau^{'})\n
&=&\dfrac{2\pi^2G}{R}i\dfrac{d}{d\tau}[I_{ij}(\tau-R)].
\end{eqnarray}
The result depends on the first time derivative of moment at retarded time.
\section{Discussion}
Our investigations show that in analysis of gravitational waves the background of de-Sitter with $K=+1$ fundamentally
 differs from the scale- free de-Sitter with $K=0$. We found the wave numbers should be discrete as already it has
 been realized that the spectrum of the Laplacian in spherical space is always discrete [17]. Another relevant feature is
 the existence of cut off on the long-wavelength of gravitational waves. This may be tested by the measurement of
 dipole and higher multi-pole moments of the CMBR anisotropy which contains information about the long-wavelength
 portion of the spectrum of energy density produced the large scale galactic structure of the universe.  These are
 sensitive to the presence of long-wavelength perturbations. The obtained eigenmodes are the fundamental
 tools of analysis of cosmic evolution of perturbations in spatially closed models. In the formation of large scale
  structure and study of anisotropies of CMBR we should use these eigenmodes to expand the perturbations. Another
 feature of the obtained eigenmodes is that they are effectively transverse in zone near to origin and at far distances, in
 contrast to flat spacetimes, this is not so. For example a wave moving in z direction its $3j$ components are not
 vanishing exactly but in near zone may be approximated to zero.  It has been shown that $h_{ij}=a^2 D_{ij}\propto
\frac{D_n(\chi,t)}{\sin^2\tau}$. This means that in a collapsing phase i.e. in the time interval $-\infty\leq t\leq 0$
corresponding to conformal time $ 0\leq\tau\leq\pi/2$, the perturbation is decaying while in the expanding phase, i.e. in
 the time interval  $0\leq t\leq+\infty$ corresponding to conformal time $\pi/2\leq \tau\leq\pi$ , it is growing. Of course
  always the smallness conditions of the perturbation with respect to the unperturbed metric holds. The amplitude of the
 gravitational wave changes with time as $h_{ij}\propto\frac{\sqrt{1+(n^2-1)\sin^2\tau}}{(n^2-1)\sin^2\tau}$.
So its growth is significant for the modes that satisfy the condition $\sqrt{n^2-1}\sin^2\tau\ll 1$, where it changes as 
$|h_{ij}|\propto\frac{1}{(n^2-1)\sin^2\tau}$. For those modes that satisfy the
 condition $\sqrt{n^2-1}\sin^2\tau\gg 1$,we have $|h_{ij}|\propto\frac{1}{\sqrt{n^2-1}\sin\tau}$ and changes are
 relatively smooth. Singularities appear in the solutions are of coordinate type, where we may see them in the unperturbed metric too. At last the obtained results are crucial
to expand any perturbation appears in different contexts of the spatially closed cosmological models in terms of them.
\newpage
\section{Appendix}
The components of the amplitude of gravitational waves moving in an arbitrary direction:
 \beqa
 {A_{\buildrel  +\over \times}}_{11}(\vec{X},\hat{n})&=&{{D}_{\buildrel  +\over \times}}_{11}(\vec{X},\hat{n}) 
\frac{\hat{n}^2_2}{1-\hat{n}^2_3}+{{D}_{\buildrel  +\over \times}}_{22}(\vec{X},\hat{n}) \frac{\hat{n}^2_1\hat{n}^2_3}{1-\hat{n}^2_3}\n
&\;&{{D}_{\buildrel  +\over \times}}_{33}(\vec{X},\hat{n})\hat{n}^2_1 +2{{D}_{\buildrel  +\over \times}}_{12}(\vec{X},\hat{n})\frac{\hat{n}_1\hat{n}_2\hat{n}_3}{1-\hat{n}^2_3}\n
&\;&2{{D}_{\buildrel  +\over \times}}_{13}(\vec{X},\hat{n})\frac{\hat{n}_1\hat{n}_2}{\sqrt{1-\hat{n}^2_3}}
+2{{D}_{\buildrel  +\over \times}}_{23}(\vec{X},\hat{n})\frac{\hat{n}^2_1\hat{n}_3}{\sqrt{1-\hat{n}^2_3}}\\
{{A}_{\buildrel  +\over \times}}_{22}(\vec{X},\hat{n})&=&{{D}_{\buildrel  +\over \times}}_{11}(\vec{X},\hat{n})\frac{\hat{n}^2_1}{1-\hat{n}^2_3}+{{D}_{\buildrel  +\over \times}}_{22}(\vec{X},\hat{n})
\frac{\hat{n}^2_2\hat{n}^2_3}{1-\hat{n}^2_3}\n
&\;&+{{D}_{\buildrel  +\over \times}}_{33}(\vec{X},\hat{n})\hat{n}^2_2-2{{D}_{\buildrel  +\over \times}}_{12}(\vec{X},\hat{n})\frac{\hat{n}_1\hat{n}_2\hat{n}_3}{1-\hat{n}^2_3}\n
&\;&-2{{D}_{\buildrel  +\over \times}}_{13}(\vec{X},\hat{n})\frac{\hat{n}_1\hat{n}_2}{\sqrt{1-\hat{n}^2_3}}
+2{{D}_{\buildrel  +\over \times}}_{23}(\vec{X},\hat{n})\frac{\hat{n}^2_2\hat{n}_3}{\sqrt{1-\hat{n}^2_3}}\\
{{A}_{\buildrel  +\over \times}}_{33}(\vec{X},\hat{n})&=&{{D}_{\buildrel  +\over \times}}_{22}(\vec{X},\hat{n})(1-\hat{n}^2_3)+{{D}_{\buildrel  +\over \times}}_{33}(\vec{X},\hat{n})\hat{n}^2_3\n
&\;&-2{{D}_{\buildrel  +\over \times}}_{23}(\vec{X},\hat{n})\hat{n}_3\sqrt{1-\hat{n}^2_3}\\
{{A}_{\buildrel  +\over \times}}_{12}(\vec{X},\hat{n})&=&-{{D}_{\buildrel  +\over \times}}_{11}(\vec{X},\hat{n})\frac{\hat{n}_1\hat{n}_2}{1-\hat{n}^2_3}+{{D}_{\buildrel  +\over \times}}_{22}(\vec{X},\hat{n})
\frac{\hat{n}_1\hat{n}_2\hat{n}^2_3}{1-\hat{n}^2_3}+{{D}_{\buildrel  +\over \times}}_{33}(\vec{X},\hat{n})
\hat{n}_1\hat{n}_2\n
&\;&+{{D}_{\buildrel  +\over \times}}_{12}(\vec{X},\hat{n})\left(\frac{\hat{n}^2_2-\hat{n}^2_1}{1-\hat{n}^2_3}\right)\hat{n}_3+{{D}_{\buildrel  +\over \times}}_{13}(\vec{X},\hat{n})\left(\frac{\hat{n}^2_2-\hat{n}^2_1}{\sqrt{1-\hat{n}^2_3}}\right)\n
&\;&+2{{D}_{\buildrel  +\over \times}}_{23}(\vec{X},\hat{n})\frac{\hat{n}_1\hat{n}_2\hat{n}_3}{\sqrt{1-\hat{n}^2_3}}\\
{{A}_{\buildrel  +\over \times}}_{13}(\vec{X},\hat{n})&=&-{{D}_{\buildrel  +\over \times}}_{22}(\vec{X},\hat{n})\hat{n}_1\hat{n}_3+{{D}_{\buildrel  +\over \times}}_{33}(\vec{X},\hat{n})\hat{n}_1\hat{n}_3-
{{D}_{\buildrel  +\over \times}}_{12}(\vec{X},\hat{n})\hat{n}_2\n
&\;&{{D}_{\buildrel  +\over \times}}_{13}(\vec{X},\hat{n})\frac{\hat{n}_2\hat{n}_3}{1-\hat{n}^2_3}
+{{D}_{\buildrel  +\over \times}}_{23}(\vec{X},\hat{n})\frac{\hat{n}_1(2\hat{n}^2_3-1)}{\sqrt{1-\hat{n}^2_3}}\\
{{A}_{\buildrel  +\over \times}}_{23}(\vec{X},\hat{n})&=&-{{D}_{\buildrel  +\over \times}}_{22}(\vec{X},\hat{n})\hat{n}_2\hat{n}_3+{{D}_{\buildrel  +\over \times}}_{33}(\vec{X},\hat{n})\hat{n}_2\hat{n}_3
+{{D}_{\buildrel  +\over \times}}_{12}(\vec{X},\hat{n})\hat{n}_1\n
&\;&-{{D}_{\buildrel  +\over \times}}_{13}(\vec{X},\hat{n})\frac{\hat{n}_1\hat{n}_3}{\sqrt{1-\hat{n}^2_3}}
+{{D}_{\buildrel  +\over \times}}_{23}(\vec{X},\hat{n})\frac{\hat{n}_2(2\hat{n}^2_3-1)}{\sqrt{1-\hat{n}^2_3}}
\eeqa
where${{D}_{\buildrel  +\over \times}}_{ij}(\vec{X},\hat{n}) $ are as follows:
\beqa
{D_+}_{11}(\vec{X},\hat{n})&=&\left[\sqrt{1-X^2}(1-\frac{(\hat{n}_2x-\hat{n}_1y)^2}{1-\hat{n}^2_3}
-(\hat{n}\cdot\vec{X})^2)\right]^{-1}~~~,\;\;\; {D_\times}_{11}(\vec{X},\hat{n})=0 \\
{D_\times}_{12}(\vec{X},\hat{n})&=&\{\sqrt{1-X^2}[1-\frac{(\hat{n}_3(\hat{n}\cdot\vec{X})-z)^2}{1-\hat{n}^2_3}-(\hat{n}\cdot\vec{X})^2]\}^{-1} , \;\;\;
{D_+}_{12}(\vec{X},\hat{n})=0\\
{D_+}_{13}(\vec{X},\hat{n})&=&\frac{(\hat{n}\cdot\vec{X})(\hat{n}_2x-\hat{n}_1y)}{\sqrt{1-X^2}\sqrt{1-\hat{n}^2_3}(1-(\hat{n}\cdot\vec{X})^2)[1-\frac{(\hat{n}_2x-\hat{n}_1y)^2}{1-\hat{n}^2_3}-(\hat{n}\cdot\vec{X})^2]}\\
{D_\times}_{13}(\vec{X},\hat{n})&=&\frac{(\hat{n}\cdot\vec{X})(\hat{n}_3(\hat{n}\cdot\vec{X})-z)}{\sqrt{1-X^2}\sqrt{1-\hat{n}^2_3}[1-\frac{(\hat{n}_3(\hat{n}\cdot\vec{X})-z)^2}{1-\hat{n}^2_3}-(\hat{n}\cdot\vec{X})^2](1-(\hat{n}\cdot\vec{X})^2)}\\
{D_+}_{22}(\vec{X},\hat{n})&=&-\{\sqrt{1-X^2}[1-\frac{(\hat{n}_3(\hat{n}\cdot\vec{X})-z)^2}{1-\hat{n}^2_3}-(\hat{n}\cdot\vec{X})^2]\}^{-1}\\
{D_\times}_{22}(\vec{X},\hat{n})&=&\frac{2(\hat{n}_2x-\hat{n}_1y)(\hat{n}_3(\hat{n}\cdot\vec{X})-z)}{(1-\hat{n}^2_3)\sqrt{1-X^2}[1-\frac{(\hat{n}_3(\hat{n}\cdot\vec{X})-z)^2}{1-\hat{n}^2_3}-(\hat{n}\cdot\vec{X})^2]^2}\\
{D_+}_{23}(\vec{X},\hat{n})&=&-\frac{(\hat{n}_3(\hat{n}\cdot\vec{X})-z)(\hat{n}\cdot\vec{X})}{\sqrt{1-X^2}(1-\hat{n}^2_3)(1-(\hat{n}\cdot\vec{X})^2)[1-\frac{(\hat{n}_3(\hat{n}\cdot\vec{X})-z)^2}{1-\hat{n}^2_3}-(\hat{n}\cdot\vec{X})^2]}\\
{D_\times}_{23}(\vec{X},\hat{n})&=&\frac{\frac{(\hat{n}_2x-\hat{n}_1y)}{\sqrt{1-\hat{n}^2_3}}(\hat{n}\cdot\vec{X})[1+\frac{(\hat{n}_3(\hat{n}\cdot\vec{X})-z)^2}{1-\hat{n}^2_3}-(\hat{n}\cdot\vec{X})^2]}{\sqrt{1-X^2}(1-(\hat{n}\cdot\vec{X})^2)[1-\frac{(\hat{n}_3(\hat{n}\cdot\vec{X})-z)^2}{1-\hat{n}^2_3}-(\hat{n}\cdot\vec{X})^2]}\\
{D_+}_{33}(\vec{X},\hat{n})&=&(\hat{n}\cdot\vec{X})^2[(\hat{n}_2x-\hat{n}_1y)^2-(\hat{n}_3(\hat{n}\cdot\vec{X})-z)^2]\{
(1-\hat{n}^2_3)\sqrt{1-X^2}(1-(\hat{n}\cdot\vec{X})^2)\n
&\;&\times[1-\frac{(\hat{n}_2x-\hat{n}_1y)^2}{(1-\hat{n}_3)^2}-(\hat{n}\cdot\vec{X})]
[1-\frac{(\hat{n}_3(\hat{n}\cdot\vec{X})-z)^2}{1-\hat{n}^2_3}-(\hat{n}\cdot\vec{X})^2]\}^{-1}\\
{D_\times}_{33}(\vec{X},\hat{n})&=&\frac{2(\hat{n}_2x-\hat{n}_1y)(\hat{n}_3(\hat{n}\cdot\vec{X})-z)(\hat{n}\cdot\vec{X})^2}{{\sqrt{1-X^2}\sqrt{1-\hat{n}^2_3}(1-(\hat{n}\cdot\vec{X})^2)[1-\frac{(\hat{n}_3(\hat{n}\cdot\vec{X})-z)^2}{1-\hat{n}^2_3}-(\hat{n}\cdot\vec{X})^2]}}
\eeqa

\end{document}